\documentclass[aps,prl,preprint,groupedaddress]{revtex4}
\usepackage{amssymb}

\input{tcilatex}

\begin{document}

\preprint{}
\title[DMC - Epidemic]{\textbf{New approach to Dynamical Monte Carlo
Methods: application to an Epidemic Model}}
\author{O.E. Aiello and Marco A.A. da Silva}
\address{\emph{Departamento de F\'{\i}sica e Qu\'{\i}mica da FCFRP,}\\
\emph{Universidade de S\~{a}o Paulo, 14040-903 Ribeir\~{a}o Preto, SP, Brazil%
}}

\begin{abstract}
A new approach to Dynamical Monte Carlo Methods is introduced to simulate
markovian processes. We apply this approach to formulate and study an
epidemic Generalized SIRS model. The results are in excellent agreement with
the forth order Runge-Kutta Method in a region of deterministic solution.\
We also demonstrate that purely local interactions reproduce a
poissonian-like process at mesoscopic level. The simulations for this case
are checked self-consistently using a stochastic version of the Euler Method.
\end{abstract}

\volumeyear{}
\volumenumber{}
\issuenumber{}
\eid{}
\date{October 29, 2001}
\received[Received text]{}
\revised[Revised text]{}
\accepted[Accepted text]{}
\published[Published text]{}
\startpage{01}
\endpage{}
\maketitle

{\Large \textit{I-Introduction}} - Monte Carlo (MC) methods have been used
mainly to equilibrium systems\cite{Binder1}, and they have broad
applications, since simple systems like hard spheres\cite{Caliri} up to
complex systems like proteins\cite{Swendsen,Cieplak}. Good reviews in
applications of MC methods to statistical physics can be seen in the
references\cite{Binder1,Binder2}. In the last decades the development of
techniques dealing with non-equilibrium systems has been increased\cite%
{Alexander}, specially those that concern with stochastic processes. Several
attempts were done \cite{Binder2}-\cite{Aiélo} to simulate real time
processes with this method. Some success was achieved within the scope of
poissonian processes \cite{Gillespie} that has been only recently properly
formalized by Fichtorn and Weinberg \cite{Fichtorn}. Another important
approach from a theoretical point of view is the waiting (or residence) time
distribution used by Prados et al.\cite{Sanchez}, whose application is
limited to simple systems, like Ising models. Some improvement in the real
time calculation was presented by Cao\cite{Cao}, but in a particular and non
rigorous way. In this letter we surmount this problem using directly the
Master Equation, ignoring thus what type of distribution we are dealing. In
this way, we also avoid the direct waiting (fine-grained) time distribution
calculation; this is substituted by the calculation of interevent
(coarse-grained) times. In our approach, the time is a dependent stochastic
variable whose distribution is constructed from the Master Equation with
appropriate transition probabilities. This gives the hierarchy of the
process. The approach is developed for a class of markovian processes with
no simultaneous events in the smallest scale considered. Thus, it is for a
restricted markovian, but more general than poissonian processes. This
method has already been applied\cite{appear} to an extensive study of the
epidemic Susceptible-Infected-Recovered-Susceptible (SIRS) systems (to
details of these epidemic systems see \cite{Aiélo} and references therein).
Here, we apply this new approach to formulate an epidemic Generalized SIRS
(GSIRS) model, and study two particular cases of it.

{\Large \textit{II-The Method}} - For discrete systems, the markovian Master
Equation is given by: 
\begin{equation}
\frac{dP_{i}(t)}{dt}=\sum_{j}w_{j\rightarrow
i}P_{j}-\sum\limits_{j}w_{i\rightarrow j}P_{i},
\label{Pauli Master Equation}
\end{equation}%
where $P_{i}$ is the probability to find the system at the state $i$ at the
time $t$, and $w_{i\rightarrow j}$ is the transition probability per unity
of time. Considering $T_{ij}$ the probability of transition from $i$ to $j$,
we may write $w_{i\rightarrow j}=\frac{T_{ij}}{\tau _{i}}$\cite{Livro},
where $\tau _{i}$ is a time constant (\textit{lifetime}) characteristic of
the state $i$.

We now start by choosing a convenient physical extensive microscopic
quantity $A_{i}$ that is time independent for each state $i$. The mean value
for this quantity at the time $t$ is given by: 
\begin{equation}
A(t)=\langle A\rangle =\sum_{i}P_{i}(t)A_{i}.  \label{Temporal Mean Value}
\end{equation}%
This equation represents a continuous physical macroscopic quantity $A(t)$.
We can differentiate both sides of the equation above, with respect to $t$.
After that, using $\left( \ref{Pauli Master Equation}\right) $, and by
defining $\Delta A_{ij}=A_{i}-A_{j}$, we get 
\begin{equation}
\frac{dA(t)}{dt}=\sum_{i}\sum_{j}w_{j\rightarrow i}P_{j}\Delta A_{ij}.
\label{Macroscopic Master Equation 2}
\end{equation}

Consider now the nearest-neighbor states $j$ of a given state $i$; if we
measure the \ ``distance'' between the states, say by the quantity $|\Delta
A_{ij}|$, such that the non-null minimum value is $|\Delta A_{ij}|=a$, we
may approach the equation$\left( \ref{Macroscopic Master Equation 2}\right) $
by:

\begin{equation}
\frac{dA(t)}{dt}=\sum_{<ij>}w_{j\rightarrow i}P_{j}a\delta _{ij},
\label{Macroscopic Master Equation 3}
\end{equation}
where the symbol \ $<ij>$ denotes a nearest-neighbour pair of states, and $%
\delta _{ij}=\Delta A_{ij}/|\Delta A_{ij}|$. Now we consider another
physical quantity $A^{\dagger }$ that is a source for the quantity $A$.
Thus, we can rewrite $\left( \ref{Macroscopic Master Equation 3}\right) $ as:

\begin{equation}
\frac{dA(t)}{dt}=\sum_{j}r_{j}^{+}P_{j}A_{j}^{\dagger
}-\sum_{j}r_{j}^{-}P_{j}A_{j},  \label{Macroscopic Master Equation 4}
\end{equation}%
where $r_{j}=<w_{j\rightarrow i}>_{i}$ are the transition probabilities per
unity of time averaged over the ensemble of the nearest-neighbour states $i$
of $j$ at some time $t$, i.e., the \textit{mesoscopic} rates. Here, ensemble
means a set of configurations accessible at a some finite (small) time
around a time $t$; in this sense we are using a time dependent ergodicity
idea\cite{Binder2}, and so generally the systems are \ non ergodic in non
equilibrium states. The \ superscripts $\ ``+"$ and $\;``-"$ mean
respectively the contributions to increasing and to decreasing the quantity $%
A(t)$. In the particular case that $r_{j}^{+}=r^{+}$ and $r_{j}^{-}=r^{-}$
are constants (or only function of the time) we have: 
\begin{equation}
\frac{dA}{dt}=r^{+}A^{\dagger }-r^{-}A,  \label{Flux Master Equation}
\end{equation}%
what is the analogous to the kinetic equation for the first order chemical
reaction $\mathcal{A}^{\dagger }\rightleftarrows \mathcal{A}$, being $%
A^{\dagger }$ and $A$ the respective concentrations of the chemical elements 
$\mathcal{A}^{\dagger }$ and $\mathcal{A}$. The equilibrium can be reached
by imposing the balance at macroscopic (or mesoscopic) level: $%
r^{+}A^{\dagger }=r^{-}A$. This follows immediately if we require the
detailed balance, but it is not necessary at all\cite{Fosdick}.

We can write the equation $\left( \ref{Macroscopic Master Equation 3}\right) 
$ in an approximated form of a discrete integral

\begin{equation}
A(t)-A(t_{0})\simeq \sum_{k=0}^{n}\sum_{<ij>}w_{j\rightarrow
i}P_{j}(t_{k})a\delta _{ij}\Delta t_{k}.
\label{Discrete integral master equation}
\end{equation}

Let now be the set of possible $w_{j\rightarrow i}$ represented by $\mathcal{%
P}_{t}=\{w_{j\rightarrow i}\}$, being the states $i$ and $j$ occurring
around a given instant $t$, and $w_{t}^{\max }=\sup \mathcal{P}_{t}$. The
phase space may be divided into $N$ parts, in such way that each part may
contain only one element of the system. Thus, each element of time in the
equation $\left( \ref{Discrete integral master equation}\right) $ may be
represented by

\begin{equation}
\Delta t_{k}=\frac{1}{w_{t_{k}}^{\max }N}.  \label{Delta T Fixed}
\end{equation}%
We can do the approach to the equation $A(t)$ considering $n=\ell N$, with $%
\ell $ sweeps over the discretized space; in the limit of $N\rightarrow
\infty $ we have the exact solution of the equation $\left( \ref{Macroscopic
Master Equation 3}\right) $ for a given initial condition.

\textit{{\large Monte Carlo Approach} -}With the considerations above the
equation $\left( \ref{Discrete integral master equation}\right) $ may be
written in the form:

\begin{equation}
A(t)-A(t_{0})=\sum_{k=0}^{\ell N}\sum_{<ij>}\left( \frac{w_{j\rightarrow i}}{%
w_{t_{k}}^{\max }}\right) \left( \frac{1}{N}\right) P_{j}(t_{k})a\delta
_{ij}.  \label{Discrete integral 2}
\end{equation}

We can create a hierarchical process choosing the probabilities of transition

\begin{equation}
T_{j\rightarrow i}^{\ast }=\frac{w_{j\rightarrow i}}{w_{t_{k}}^{\max }},
\label{Transition probalities}
\end{equation}
that reproduce the correct frequencies of events at each time $t_{k}$ to
solve $\left( \ref{Discrete integral 2}\right) $. This hierarchy have subtle
differences with an earlier hierarchy introduced by Fichtorn et al\cite%
{Fichtorn}: first in that work (mesoscopic) rates were required, while here
we primarily use \textit{transition probability per unity of time. }Second,
they used a global maximum to the rates, while here we use a more local
maximum; in recent work\cite{Aiélo} this was done without a rigorous proof,
based only in the detailed balance principle applied to a specific case. To
carry out the MC procedure, an element is selected randomly with a
probability $\frac{1}{N}$, and thus a transition is tried with probability
given by $\left( \ref{Transition probalities}\right) $. The space is swept $%
\ell $ times, with the increment of time in each MC step (one MC step here,
means a single try to change the state of one element of the system) given
by $\left( \ref{Delta T Fixed}\right) $ up to reach a time $t$. Starting
from the same initial conditions for the physical quantities, the process
may be repeated, and we can get the average quantity $A(t)$ at each instant $%
t$. We must emphasize that the probabilities $P_{j}$ are generated by this
process. As a given state is chosen with its correct probability in a given
time, an ideal MC procedure leads to

\begin{equation}
A(t)-A(t_{0})=\sum_{k=0}^{\ell N}(\left\langle r^{+}A^{\dagger
}\right\rangle _{j_{k}}-\left\langle r^{-}A\right\rangle _{j_{k}})\left( 
\frac{1}{w_{t_{k}}^{\max }N}\right) ,
\end{equation}%
where $\ $the \ averages are taken over the ensemble of the states $j_{k}$
at each instant $t_{k}$. This is just an approach to the integration result
of the equation $\left( \ref{Macroscopic Master Equation 4}\right) $.

We need to observe some important points: first, generally different runs
give different time $t_{k}$ results at the same MC step $k$, and the sample
averages may be done by linear interpolating or extrapolating the data set,
in each MC realization, to do them at the same point of the time. Second, in
one complete sweep around a time $t_{k}$, the value $w_{t_{k}}^{\max }$ must
be approximately constant in order do not change the hierarchy and so the
result. Third, as the configurations do not change drastically in few steps,
the microscopic transitions reproduce the mesoscopic result.

Another approach consists in estimating the interevent times by the
following rule

\begin{equation}
\Delta t_{k}^{e}=\frac{f_{e}^{k}a}{r_{j_{k}}^{e}A_{j_{k}}^{e}},
\label{Interevent time}
\end{equation}%
where $r_{j_{k}}^{e}=r_{j_{k}}^{+}$ and $A_{j_{k}}^{e}=A_{j_{k}}^{\dagger }$%
, or, $r_{j_{k}}^{e}=r_{j_{k}}^{-}$ and $A_{j_{k}}^{e}=A_{j_{k}}$ depending
on, respectively, if the outcome of the experiment increase or decrease the
quantity $A$. The quantity $f_{e}^{k}$ is an arbitrary $e$-event dependent
factor that must obey the relationship $\sum\limits_{e}f_{e}^{k}=1$, for
each time $t_{k}$. We emphasize that the time given by $\left( \ref%
{Interevent time}\right) $ represents the average waiting time to
transitions from a given state $j_{k}$ to any neighbor state $i$; if the
microscopic state remains unchanged, the time does not evolve. It can be
shown that this procedure leads to the same result as using $\left( \ref%
{Delta T Fixed}\right) $ at each MC step observing that

\begin{equation}
\Delta t_{k}=\sum_{e}\sum_{i}\left( \frac{w_{j_{k}\rightarrow i}}{%
w_{t_{k}}^{\max }}\right) \left( \frac{1}{N}\right) \Delta t_{k}^{e}.
\label{Delta T fixed 2}
\end{equation}%
As $r_{j_{k}}^{e}A_{j_{k}}^{e}=a\sum\limits_{i}w_{j_{k}\rightarrow i}$,
using the equation $\left( \ref{Interevent time}\right) $ and the
normalization condition to $f_{e}^{k}$ in $\left( \ref{Delta T fixed 2}%
\right) $, we obtain the expression $\left( \ref{Delta T Fixed}\right) $. In
particular, if we choose $f_{e}^{k}=0$, for most events $e$, except some $%
e=s $, we have $f_{s}^{k}=1$, so, with this condition, the interevent time
has the meaning of the waiting time between type-$s$ events. Based on this
and in the fact that at the equilibrium the relative frequencies of
occurrence of events are all equal, we may define $f_{e}^{k}\equiv n_{e}^{k}/%
\mathcal{N}_{k}$,\ where $n_{e}^{k}$ is the number of $e-$events, and $%
\mathcal{N}_{k}=\sum_{e}n_{e}^{k}$ is the total number of events, in a time
interval (arbitrary) near to some time $t_{k}$.

\textit{\Large III-GSIRS model} - Based on $\left( \ref{Macroscopic Master
Equation 4}\right) $, we formulated the GSIRS model through the following
set of differential equations and inter-classes rates:

\begin{eqnarray}
\frac{dS}{dt} &=&\sum_{j}^{j}r_{R\rightarrow
S}^{j}P_{j}R_{j}-\sum_{j}r_{S\rightarrow I}^{j}P_{j}S_{j},  \label{DSDT} \\
\frac{dI}{dt} &=&\sum_{j}r_{S\rightarrow
I}^{j}P_{j}S_{j}-\sum_{j}r_{I\rightarrow R}^{j}P_{j}I_{j},  \label{DIDT} \\
\frac{dR}{dt} &=&\sum_{j}r_{I\rightarrow
R}^{j}P_{j}I_{j}-\sum_{j}r_{R\rightarrow S}^{j}P_{j}R_{j},  \label{DRDT}
\end{eqnarray}%
where $S$, $I,$ and $R$ are the populational classes, respectively, of the
number of individuals in the susceptible, infective and recovered classes.
Being the mesoscopic rates $r_{S\rightarrow I}^{j}$ , $r_{I\rightarrow
R}^{j} $ and $r_{R\rightarrow S}^{j}$, for each state $j$, respectively,
from $S\rightarrow I$, $I\rightarrow R$ and $R\rightarrow S$. Note that we
meant that, for example, if $A=I$, then $A^{\dagger }=S$ in the equation $%
\left( \ref{Macroscopic Master Equation 4}\right) $. The conservation law
with the total number of individuals $N=S(t)+I(t)+R(t)$ is satisfied. In
particular, a model commonly used\cite{Aiélo,Haas} give $w_{R\rightarrow
S}=m,w_{S\rightarrow I}=\Gamma \,\frac{b}{N^{\mu }}S^{\mu -1}I+\Lambda
\,[1-(1-p_{0})^{n}],$ and $w_{I\rightarrow R}=q$ to the transition
probabilities per unity of time. We must observe that the mesoscopic rates
are resulting from local (``instantaneous'') averages of the respective
transition probabilities per unity of time. For practical purposes the
individuals are distributed on a square lattice of $N=M\times M$ sites. All
the individuals at the lattice boundary have their states fixed at
susceptible state.

\textit{\Large IV-Results and Conclusions} - We set the lattice size to $%
M=200$. This size was sufficient to get good results compared with the
continuum limit when only global interactions ($\Lambda =0$) are considered.
The initial condition for the system is set up by $I_{0}=2000$ infectives
being randomly distributed on the lattice and the remaining sites being
occupied by $S_{0}=N-I_{0}$ susceptibles, so $R_{0}=0$.

We consider here two particular cases of the system defined by $\left( \ref%
{DSDT}-\ref{DRDT}\right) $. First, we set $\Lambda =0$, and the other model
parameters as $q=0.2,$ $b=0.8,$ $m=0.01$ and $\mu =2$. The non-minimum
value, to the differences $\Delta S,$ $\Delta I$ and $\Delta R$ , used in$%
\left( \ref{Interevent time}\right) $ is $a=|\Delta I|=|\Delta S|=|\Delta
R|=1$. Figure 1 shows the temporal evolution of $I(t).$ Continuous lines
represent numerical (fourth-order Runge-Kutta) \emph{checking solutions} for
the set of differential equations $\left( \ref{DSDT}-\ref{DRDT}\right) $,
and open circles correspond to the MC simulations. The accuracy of the
deterministic solution (Runge-Kutta) was estimated as less than $0.1\%$ (see
ref.\cite{Aiélo}). \emph{\ } Results to the system\ far from equilibrium
showed that the interevent times given by $\left( \ref{Interevent time}%
\right) $ have poissonian-like distribution (see inset in figure 1) as
expected\cite{Aiélo}. At the equilibrium, the present method leads to
converge the distributions of interevent times to delta distributions,
because the values to the rates and other physical quantities converge to
constant values. A total of $4\times 10^{6}$ steps, corresponding to $%
3,5\times 10^{5}$ configurations, was generated by the MC procedure, leading
to a total real time of approximately $500$ days. The total number of
configurations used to get the interevent times distribution was about $%
8\times 10^{4}$, what corresponds to approximately $60$ days. Second, we set 
$\Gamma =0$ and $m=0$, i.e., a SIR system with purely local variables. The
variable $n$ \ is an integer ranging from $n=0$ up to $8$, since the first
and second nearest infected neighbors are indistinguishably considered for
each susceptible. To this case we use again the expression $\left( \ref%
{Interevent time}\right) $, but the rates $r_{S\rightarrow I}$\ are obtained
by averaging the individual probabilities to the configurations in every
successful event. This may coast some simulation time. A good optimization
for an approximation to the exact average is done by drawing randomly
susceptibles ($1000$ here was sufficient) for each configuration reached and
doing a sample mean with the site transition probabilities per unit of time $%
w_{S\rightarrow I}$. It must be observed that this type of average is
equivalent to let the system advance some small time and take an average
over the sample. As the system configurations do not change much around some
time $t_{k}$, the small time average corresponds to an average in an
instantaneous time. To see the self-consistency of the approach, we
integrate numerically $\left( \ref{DSDT}-\ref{DRDT}\right) $ given constant
(or piecewise constant) time step as in $\left( \ref{Discrete integral
master equation}\right) $ by choosing the maximum local transition
probability per unit of time. This maximum is in fact actualized at each MC
step, when necessary, using a table. When a transition changes a state of an
individual that changes the maximum, the table is updated. The quantities $%
S,I$ and $R$ are calculated with iterations; the rates are chosen randomly
by the MC procedure, and thus we use the Euler Method procedure to solve
first order differential equations. Experiments using poissonian
distributions\cite{Fichtorn} to obtain the interevent times showed that the
processes are poissonian-like to all ranges of $p_{0}$, being so,
unnecessary the hypothesis of low $p_{0}$ (``weak interaction'') as done by
Aiello et al\cite{Aiélo}. To illustrate, we show in the Figure 2 the results
to $p_{0}=0.8$. We compare, also, in figure 2 the iterative method with the
MC technique described above (restricted markovian method), estimating the
interevent time by $\left( \ref{Interevent time}\right) $. \ The total
number of configurations used in the MC procedure was about $4\times 10^{4}$
what gives approximately $10$ days. The results are in excellent agreement
among them. For both cases (Figures 1 and 2), results with respect to the MC
simulations correspond to an average of $20$ independent trajectories. The
typical MC data errors are in the interval $0.1$-$1.0\%$, so most of the
error bars are smaller than the symbols in the figures.

We believe that the class of epidemic SIRS models studied here are
poissonian-like in the mesoscopic scale because of two factors. First, the
approach itself implies that no two or more events occur in a short scale of
time. Second, the mesoscopic rates are slowly varying with the time,
resembling the independence between events. So, the two conditions for a
poissonian process were met. We emphasize that low correlations between
events are not required. It is necessary that the results for independent
runs be uncorrelated, so we can use the averages obtained for each time t to
represent properly the physical quantities of the process. To do this we
need a local equilibrium hypothesis, what may be at first glance
restrictive, however we may even reduce the time observation sufficiently
such that the system does not have time to leave some metastable states. So,
we can average it there. In the practice of the simulation this is done by
increasing the number of observations, i.e., the number of time experiments.
In forthcoming works we expect to generalize still more the method,
including up to non-markovian processes.

The authors gratefully acknowledges funding support from FAPESP Grant n.
00/11635-7 and 97/03575-0. The authors would also like to thank Drs. F.L.B.
da Silva and A. Caliri for many stimulating discussions and suggestions.

\newpage

\begin{center}
{\Large Figure Captions}
\end{center}

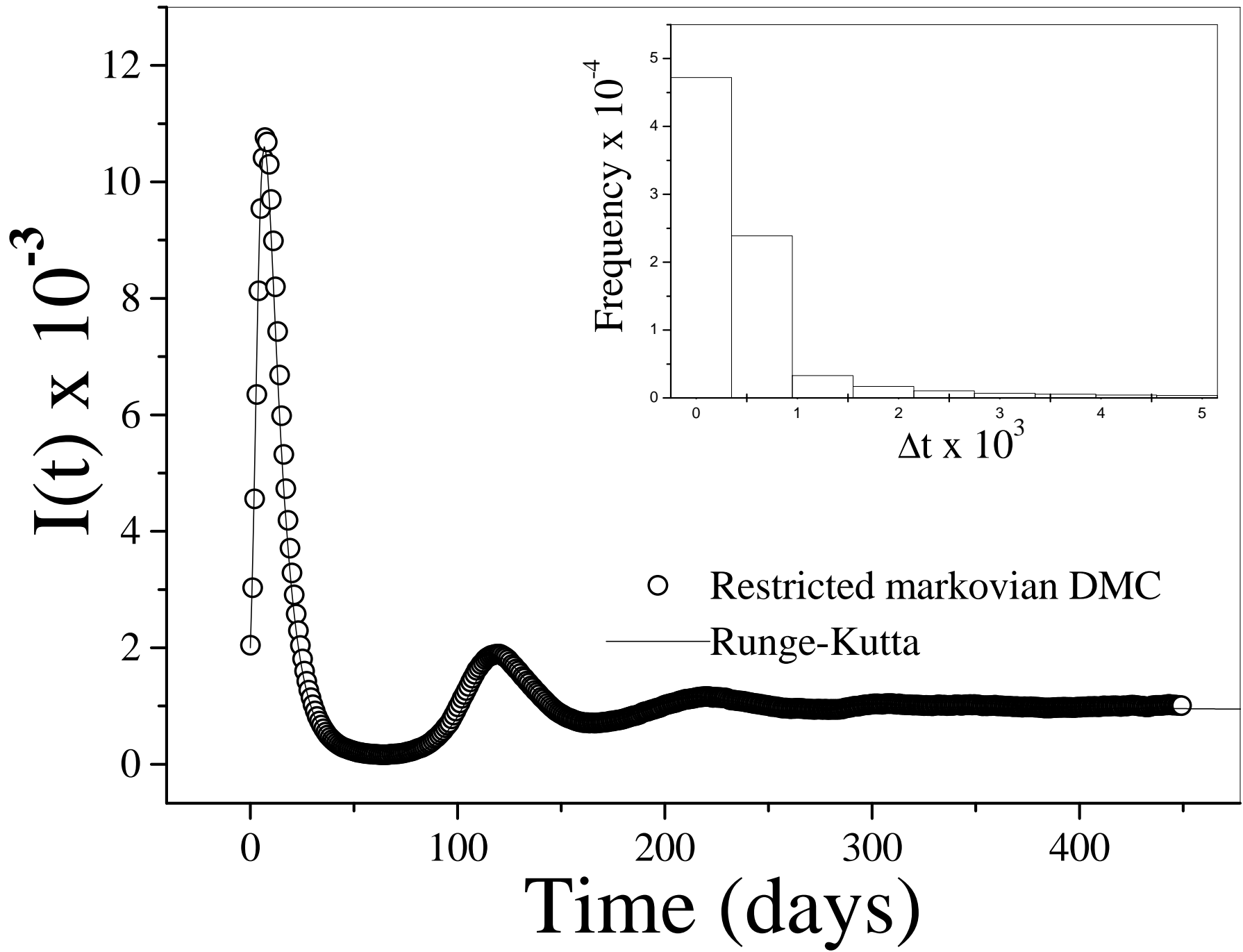
FIG. 1. Infected numbers I(t) vs Time. Continuos line: numerical forth-order
Runge-Kutta solution. Open circles: restricted markovian DMC simulation.
Inset: shows the behavior of the interevent time $\Delta t$ distribution.

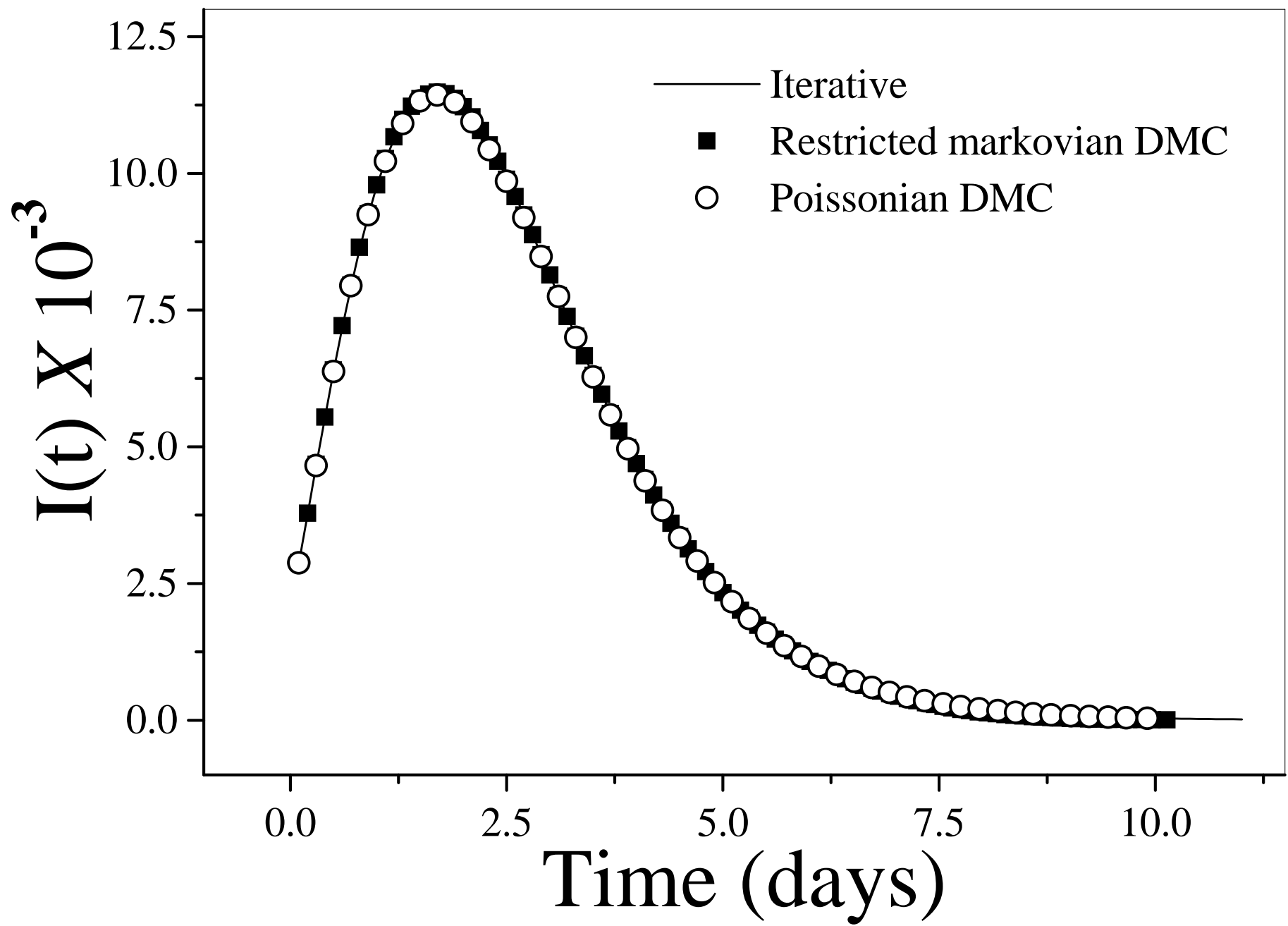
FIG. 2. Infected numbers I(t) vs Time. Continuos line: Iterative stochastic
Euler Method solution. Squares: restricted markovian DMC simulation. Open
circles: poissonian DMC simulation.


\bigskip

\newpage

\begin{thebibliography}{99}
\bibitem{Binder1} K.\ Binder, \emph{Monte Carlo Method in Statistical Physics%
} (Spriger-Verlag, Berlin,\ 1986).

\bibitem{Caliri} A. Caliri, M. A. A. da Silva, and B. J. Mokross, J. Chem.
Phys. \textbf{91}, 6328 (1989).

\bibitem{Swendsen} D. Bouzida, S. Kumar, and R. H. Swendsen, Phys. Rev. A 
\textbf{45},8894 (1992).

\bibitem{Cieplak} M. Cieplak, M. Henkel, J. Karbowski, and J.R. Banavar,
Phys. Rev. Lett. \textbf{80}, 3654 (1998).

\bibitem{Binder2} K.\ Binder, Rep. Prog. Phys. \textbf{60}, 487 (1997).

\bibitem{Alexander} F.J. Alexander, A. L. Garcia, and B. J. Alder, in \emph{%
25 Years of Non-Equilibrium Statistical Mechanics}, edited by J. J. Brey et
al (Springer-Verlag, Barcelona, Spain, 1994).

\bibitem{Sanchez} A. Prados, J.J. Brey, and B. S\'{a}nchez-Rey, Journal of
Statistical Physics \textbf{89}, 709 (1997).

\bibitem{Gillespie} D. T. Gillespie, J. Comp. Phys. \textbf{22}, 403 (1976).

\bibitem{Fichtorn} K. A. Fichtorn and W. H. Weinberg, J. Chem. Phys. \textbf{%
95}, 1090 (1991).

\bibitem{Cao} Pei-Lin Cao, Phys. Rev. Lett. \textbf{73}, 2595 (1994).

\bibitem{Aiélo} O.E. Aiello, V.J. Haas, A. Caliri, and M. A. A. Silva,
Physica A. \textbf{282}, 546\ (2000).

\bibitem{Livro} P.G. Hoel, S.C. Port, and C.J. Stone, \emph{Introduction to
Stochastic Processes} (Waveland Press, Inc., Prospect Heights, Illinois,
1987).

\bibitem{Fosdick} L. D. Fosdick, in\ \emph{Methods Comp. Phys.}, edited by
B. Alder, S. Fernback and M. Rotenberg, Vol. 1 (Academic Press, 1963), p.
245.

\bibitem{appear} O.E. Aiello and M. A. A. Silva (to be published).

\bibitem{Haas} V.J. Haas, A. Caliri, and M.A.A. da Silva, J. of Biol. Phys., 
\textbf{25}, 309 (1999).
\end{thebibliography}
\end{document}